Table of Contents Graphic

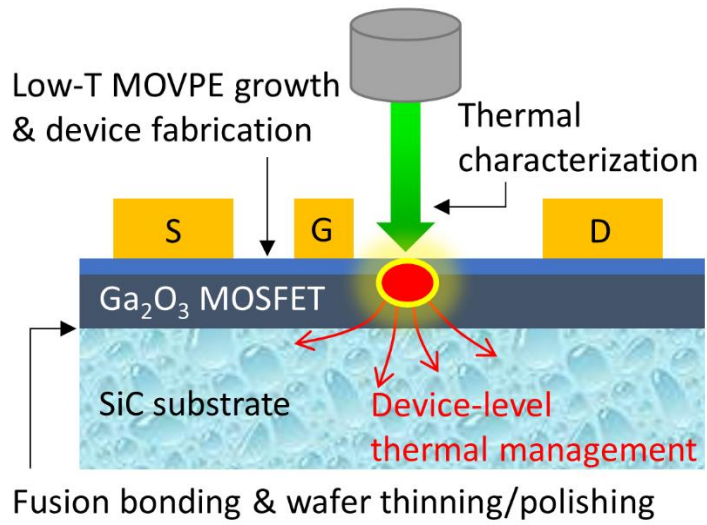

# Ultra-Wide Bandgap Ga$_2$O$_3$-on-SiC MOSFETs

Yiwen Song[1,☉], Arkka Bhattacharyya[6,☉], Anwarul Karim[1], Daniel Shoemaker[1], Hsien-Lien Huang[3], Saurav Roy[6], Craig McGray[4], Jacob H. Leach[5], Jinwoo Hwang[3], Sriram Krishnamoorthy*[,6], Sukwon Choi*[,1]

([☉] contributed equally to this work)

[1] Department of Mechanical Engineering, The Pennsylvania State University, University Park, Pennsylvania 16802, USA

[2] Department of Electrical and Computer Engineering, University of Utah, Salt Lake City, Utah, 84112, USA

[3] Department of Materials Science and Engineering, The Ohio State University, Columbus, Ohio, 43210 USA

[4] Modern Microsystems, Gaithersburg, Maryland 20878, USA

[5] Kyma Technologies, Inc., Raleigh, North Carolina 27617, USA

[6] Materials Department, University of California, Santa Barbara, Santa Barbara, California 93106, United States

**ABSTRACT:** Ultra-wide bandgap semiconductor devices based on β-phase gallium oxide (Ga$_2$O$_3$) offer the potential to achieve higher switching performance, efficiency, and lower manufacturing cost than today's wide bandgap power electronics. However, the most critical challenge to the commercialization of Ga$_2$O$_3$ electronics is overheating, which impacts the device performance and reliability. We fabricated a Ga$_2$O$_3$/4H-SiC composite wafer using a fusion-bonding method. A low temperature (≤ 600°C) epitaxy and device processing scheme was developed to fabricate MOSFETs on the composite wafer. The low-temperature grown epitaxial Ga$_2$O$_3$ devices deliver high thermal performance (56% reduction in channel temperature) and a power figure of merit of (~300 MW/cm$^2$, which is highest among heterogeneously integrated Ga$_2$O$_3$ devices reported to date. Simulations calibrated based on thermal characterization results of the Ga$_2$O$_3$-on-SiC MOSFET reveal that a Ga$_2$O$_3$/diamond composite wafer with a reduced Ga$_2$O$_3$ thickness (~1 μm) and thinner bonding interlayer (<10 nm) can reduce the device thermal impedance to a level lower than today's GaN-on-SiC power switches.

*Keywords: Composite substrate, gallium oxide (Ga$_2$O$_3$), power electronics, Raman spectroscopy, steady-state thermoreflectance, thermal management, ultra-wide bandgap (UWBG) semiconductor devices.*

## 1. INTRODUCTION

β-phase gallium oxide (Ga$_2$O$_3$) is an ultra-wide bandgap (UWBG) semiconductor ($E_G$~4.8 eV), which promises significant improvements in the performance and manufacturing cost over today's commercial wide bandgap (WBG) power electronic devices based on GaN and SiC.[1] During the past half-decade, significant progress has been made in the Ga$_2$O$_3$ bulk material synthesis (i.e., only Ga$_2$O$_3$ offers melt-grown single crystal substrates like Si wafers, among today's WBG and UWBG semiconductors), epitaxial growth, doping, and the development of homoepitaxial device architectures.[1–3] UWBG Ga$_2$O$_3$ electronics give promise to allow designers to use fewer devices and smaller passive components in power electronics circuits.[4] Power conversion systems for electric vehicles and charging stations, renewable energy sources, and smart grids will benefit from the Ga$_2$O$_3$ device technologies.

However, device overheating has become one of the most critical bottlenecks to the commercialization of Ga$_2$O$_3$ device technologies.[5] In fact, no Ga$_2$O$_3$ device reported to date has achieved the performance expected by the outstanding electronic properties because a thermally limited technological plateau has been reached. Ga$_2$O$_3$ possesses a poor anisotropic thermal conductivity (11-27 W/mK)[6,7], which is an order of magnitude lower than those for GaN (~150 W/mK)[8,9] and SiC (~400 W/mK)[10,11]. It has been experimentally reported that single-finger Ga$_2$O$_3$ metal-oxide-semiconductor field effect transistors (MOSFETs)[12] and modulation-doped FETs (MODFETs)[13,14] exhibit a ~6× higher channel temperature rise than commercial GaN high electron mobility transistors (HEMTs) under identical power dissipation levels. Moreover, recent computational work[15] indicates that self-heating will be significantly aggravated in practical multi-finger devices due to the thermal cross-talk[16] among adjacent current channels. Specifically, it has been predicted that the channel temperature rise of a six-finger Ga$_2$O$_3$ MOSFET would be another 4× higher than that for a single finger Ga$_2$O$_3$ device. Such aggravated self-heating in multi-channel Ga$_2$O$_3$ FinFETs as compared to single-fin devices has been experimentally demonstrated.[17] This signifies the importance of minimizing the junction-to-package thermal resistance of Ga$_2$O$_3$ devices.

Efforts to counter the overheating at the package/system-level not only increase the system size and weight but also have proven to be ineffective in cooling ultra-high power density WBG/UWBG devices.[18] Therefore, the electro-thermal co-design of novel device architectures that can simultaneously achieve the lowest thermal resistance and highest electrical performance is essential to enable the commercialization of UWBG Ga$_2$O$_3$ device technologies.[18]

In our previous work[15], a Ga$_2$O$_3$/4H-SiC composite wafer was created by taking advantage of a fusion bonding process. The thermal conductivity of the Ga$_2$O$_3$ layer and the thermal boundary resistance at the Ga$_2$O$_3$/SiC interface were characterized via a steady-state thermoreflectance technique. Scanning transmission electron microscopy and energy dispersive X-ray spectroscopy were used to study the interface quality and chemistry. In this work, Ga$_2$O$_3$ MOSFETs were fabricated on the composite substrate using low temperature ($\leq$ 600°C) metalorganic vapor-phase epitaxy, that allowed the first realization of "homoepitaxial" Ga$_2$O$_3$ MOSFETs on a composite substrate. This growth technique is necessary to prevent interface failure of the composite substrate due to mismatch of the coefficients of thermal expansion. Electrical testing was performed to determine the device output characteristics and breakdown voltages. The enhanced device thermal impedance achieved by integration with the composite wafer was assessed by using nanoparticle-assisted Raman thermometry. A design optimization study has been conducted with an aim to reduce the device thermal impedance of Ga$_2$O$_3$ transistors below that for a commercial GaN power switch under high power and frequency switching operation. The outcomes of this work provide guidelines to surpass the ultimate thermal limit of the (laterally configured) UWBG device technology.

## 2. RESULTS AND DISCUSSION

### 2.1. COMPOSITE WAFER FABRICATION

In our previous work[15], a Ga$_2$O$_3$/4H-SiC composite wafer was fabricated using a fusion bonding process. Here we provide a summary of the fabrication process and structure of the composite wafer. 15 nm of SiN$_x$ was coated on both a (010)-oriented Fe-doped 25 mm-diameter Ga$_2$O$_3$ wafer (which is the preferred orientation to achieve higher thermal performance)[13,19] and a 50 mm-diameter 4H-SiC wafer. To initiate fusion bonding[20,21], the wafer surfaces were activated in oxygen plasma and joined at room temperature. After this, the integrated material stack was cured at 215°C in a N$_2$ convection oven. The bonded Ga$_2$O$_3$ wafer was thinned down using a series of lapping plates and a diamond abrasive (9 µm, 3 µm, and 0.25 µm diamond grit size). A silica-based chemical-mechanical polishing (CMP) process was used to achieve an epitaxial-ready surface. A critical advantage of the wafer bonding approach over heteroepitaxy directly on SiC[22] is that it allows subsequent homoepitaxial growth of highest quality crystalline films without threading dislocations, because the starting material is a highest quality melt-grown Ga$_2$O$_3$ substrate. The final thickness of the Ga$_2$O$_3$ was measured via cross-sectional scanning electron microscopy (SEM) of a specimen prepared by focused ion beam (FIB) milling. Scanning transmission electron microscopy (STEM) imaging and energy-dispersive X-ray spectroscopy (EDX) mapping were respectively used to evaluate the quality and chemistry of the Ga$_2$O$_3$/SiC interface. As shown in **Figure 1**, a 10 nm thick SiO$_x$ layer was unintentionally formed between the SiN$_x$ bonding layers, contributing to the effective thermal boundary resistance at this interface.[23]

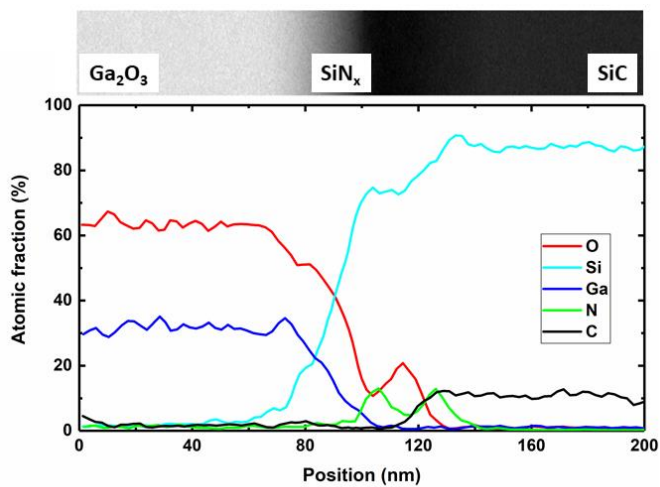

**Figure 1**: High angle annular dark field STEM image near the Ga$_2$O$_3$/4H-SiC interface and EDX line map across the Ga$_2$O$_3$/4H-SiC interface.

## 2.2. THERMO-PHYSICAL PROPERTY MEASUREMENT

Here we provide a summary of the results for thermal characterization performed on a composite wafer with a $Ga_2O_3$ thickness of 6.5 μm in our previous work[15]. The total thermal resistance of the composite wafer is dictated by the thermal conductivities of the 4H-SiC wafer and the $Ga_2O_3$ layer, as well as the thermal boundary resistance (TBR) at the $Ga_2O_3$/4H-SiC interface. The thermal conductivity of the 4H-SiC substrate was measured by a steady-state thermoreflectance (SSTR) technique[24], before integrating with a $Ga_2O_3$ substrate. The uncertainty in the SSTR measurement results is relatively high; however, the mean value of the measured SiC thermal conductivity agrees well with those reported by Wei et al.[10] The post-integrated $Ga_2O_3$ film's thermal conductivity was measured by adjusting the probing depth of the SSTR pump laser (which is dictated by the radius of the pump laser[24]) to a level smaller than the thickness of the $Ga_2O_3$ layer. Finally, the TBR across the $Ga_2O_3$/SiC interface was extracted by performing SSTR measurement of the composite wafer with a large pump laser diameter that would allow probing the thermal resistance of a volume that reaches below the surface of the 4H-SiC substrate. A TBR of 47.1 $m^2K/GW$ was derived from this differential SSTR measurement procedure, based on the pre-determined thermal conductivities of the 4H-SiC substrate and the $Ga_2O_3$ layer (**Table 1**). Time domain thermoreflectance (TDTR; used to measure the cross-plane thermal conductivity of the $Ga_2O_3$ film)[25,26] and frequency domain thermoreflectance (FDTR; used to measure the TBR at the $Ga_2O_3$/SiC interface)[27,28] were used to confirm the SSTR-based TBR results by characterizing a sister-sample with the $Ga_2O_3$ layer further thinned down to ~2 μm. The TBR extracted from this approach ($42.8^{+20.6}_{-10.5}$ $m^2K/GW$) agrees well with the TBR value extracted by SSTR listed in **Table 1**. It should be noted that the experimental determination of the TBR is challenging due to the low measurement sensitivity;[29] therefore, analytical modeling was also used to confirm the mean value of the TBR measured by SSTR. Details of this combined approach using SSTR, TDTR, and FDTR methods as well as analytical modeling to estimate the effective TBR at the $Ga_2O_3$/SiC interface (including the measurement sensitivity analyses) can be found in our previous work.[15]

Table 1 Thermal properties of the composite substrate

| | |
|---|---|
| SiC Thermal Conductivity | 306.4 ± 79.5 W/mK |
| $Ga_2O_3$ Thermal Conductivity | 19.4 ± 3.03 W/mK |
| Thermal Boundary Resistance | 47.1 $m^2K/GW$ |

The TBR of the composite wafer is lower than a value (60 $m^2K/GW$) that was shown to be necessary to reduce the junction-to-package thermal resistance of a $Ga_2O_3$-on-SiC device below that of a commercial GaN-on-Si power switch.[13] However, it is 3× higher than a value reported for a $Ga_2O_3$/SiC interface joined with a 30 nm $Al_2O_3$ bonding layer.[29] The relatively high TBR is caused by the low thermal conductivities of the $SiN_x$ bonding layer (1.9 W/mK)[30] and the unintentionally formed $SiO_x$ (1.1 W/mK)[31] interlayer. Higher thermal performance of the composite wafer can be achieved by optimizing the bonding process, i.e., using a higher thermal conductivity bonding material and minimizing its thickness, as well as eliminating the formation of a $SiO_x$ interlayer.

In this work, as mentioned in the following subsection, a composite wafer with a $Ga_2O_3$ thickness of ~34 μm was used for subsequent device processing. This composite wafer was simultaneously fabricated with the composite wafer (with a $Ga_2O_3$ thickness of 6.5 μm) reported in our previous work[15]. Therefore, we assume an identical TBR for both composite wafers.

## 2.3. DEVICE FABRICATION

To maintain the structural integrity of a composite substrate, it is of critical importance to limit the maximum temperature that occurs during the multiple processing steps associated with device fabrication. For this reason, a low temperature device processing scheme was developed that keeps the maximum temperature of the entire process below 600°C. It should be noted that current growth techniques for other electronic grade WBG and UWBG materials systems such as GaN, SiC, and AlGaN do not allow this.[32–34] Device fabrication began with the epitaxial growth of a Si-doped channel. The composite substrate was solvent cleaned in sonication baths of acetone, isopropyl alcohol and deionized (DI) water. Next, the substrate was dipped in a hydrofluoric acid solution for 15 minutes and then cleaned using DI water. After cleaning, the sample was loaded into a metalorganic vapor-phase epitaxy (MOVPE) reactor and a (010) oriented Si-doped $Ga_2O_3$ epilayer (~400 nm thick) was grown at 600°C.[35] An Agnitron Agilis vertical quartz tube MOVPE reactor was used with triethylgallium (TEGa) and $O_2$ as the precursor gases, argon as the carrier gas, and diluted silane for doping. From Hall measurements, the room temperature channel sheet charge and mobility were found to be $1.3\times10^{13}$ cm$^{-2}$ and 94 cm$^2$/Vs, respectively. Due to the sufficient adatom diffusion lengths at this growth condition (i.e., temperature and molar gas flow fluxes), an atomically smooth surface morphology (RMS roughness of ~0.5 ± 0.1 nm) was maintained and single crystal films with high crystalline quality and transport properties were realized. The low temperature MOVPE growth process helps minimize the potential risk of debonding of the composite wafer due to the thermal expansion mismatch between the Ga$_2$O$_3$ and 4H-SiC materials.[15] A cross-sectional schematic of a Ga$_2$O$_3$ metal-oxide-semiconductor field-effect transistor (MOSFET) fabricated on the composite wafer is shown in **Figure 2 (a)**. Device processing started with mesa isolation using a patterned Ni/SiO$_2$ hard mask and directional dry etching, i.e., inductively coupled plasma - reactive ion etching (ICP-RIE) SF$_6$-Ar (600W ICP, 150 RF powers – 45 nm/min etch rate for Ga$_2$O$_3$).[36,37] This was followed by source-drain region patterning using the same Ni/SiO$_2$ patterning process and contact region recessing using a low power ICP-RIE SF$_6$-Ar (150W ICP, 50 RF powers – 1.5 nm/min etch rate for Ga$_2$O$_3$).[38] After selectively wet etching Ni, the sample with the patterned SiO$_2$ mask was loaded into the MOVPE reactor for ohmic contact regrowth. A heavily Si-doped n+ (estimated $1.4\times10^{20}$ cm$^{-3}$) Ga$_2$O$_3$ layer was grown at 600°C with an approximate thickness of 100 nm.[37] The sample was then cleaned in an HF solution and the regrowth mask including regrown Ga$_2$O$_3$ was selectively removed from all regions except the source-drain regions. This was followed by ohmic metal evaporation of Ti/Au/Ni (20/100/50 nms) on the n+ regions by photolithography and lift-off. The contacts were then annealed in a rapid thermal processing (RTP) furnace at 450°C for 1.5 mins in a N$_2$ ambient. A thermal atomic layer deposition (ALD) grown Al$_2$O$_3$ layer (at 250°C) with a target thickness of 25 nm was blanket deposited to form the gate insulator. Then, a Ni/Au/Ni (30/100/30 nm) metal stack was evaporated to form the gate electrode. Finally, the Al$_2$O$_3$ over the source-drain metal pad regions was removed using a photoresist mask and CF$_4$-O$_2$-N$_2$ ICP-RIE dry etching. It is to be noted that (apart from benefitting from the low temperature processes), the composite substrate shows strong ruggedness against standard device processing steps (such as ultrasonication in solvents, acid cleaning, wet and dry etching, and dielectric/metal deposition), as well as patterning processes. This confirms its compatibility with standard device processing methods used to fabricate homoepitaxial devices on bulk Ga$_2$O$_3$ wafers.

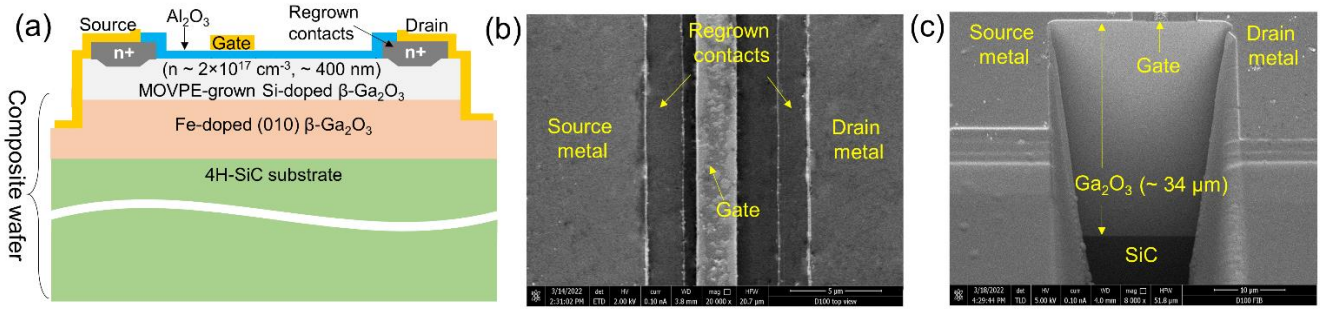

**Figure 2**: **(a)** A cross-sectional schematic of a Ga$_2$O$_3$ MOSFET fabricated on the composite substrate. **(b)** Plan-view SEM image of a final device structure. **(c)** Cross-sectional SEM image of the same device showing the thickness of Ga$_2$O$_3$ layer.

The device dimensions were verified by top-side SEM imaging (**Figure 2 (b)**). The $L_{GS}$ and $L_G$ were fixed at ~0.7 nm and ~2.1 μm, respectively, while the $L_{GD}$ was varied from ~2.5 to 55 μm. The thickness of the Ga$_2$O$_3$ layer was determined to be ~34 μm using cross-sectional SEM imaging (**Figure 2 (c)**). From transfer length method (TLM) measurements, the contact resistance to the channel was ~1.6 ± 0.2 Ω.mm. **Figures 3 (a)** and **(b)** show the direct current (DC) output and transfer curves, respectively, for a device with $L_{GD}$ ~ 2.5 μm. A device with $L_{GD}$ ~ 2.5 μm exhibits a drain current of ~100 mA/mm at a drain-source voltage of 8 V and gate bias of 0 V. The ON resistance from the linear region of the output curve is ~ 65 Ω·mm. From the transfer curve, the device shows clear pinch-off characteristics and the ON/OFF ratio is ~10$^8$. The device showed a large threshold voltage of -50 V, most likely due to the presence of a remnant active parasitic channel at the epilayer/Ga$_2$O$_3$ (of the composite substrate) interface. **Figure 3 (c)** shows the channel charge profile that is extracted from capacitance-voltage (C-V) measurements. A clear charge peak can be seen at the epilayer-substrate interface, potentially originating from the polishing step used to thin down the bonded Ga$_2$O$_3$ wafer. The magnitude of this parasitic charge was spatially nonuniform across the composite wafer. This is also revealed by the nonuniform threshold (or pinch-off) voltage of transistors fabricated on the composite substrate, which varied from -40 V to -85 V. This observation indicates a parasitic charge of $3 – 8\times10^{12}$ cm$^{-2}$ at the epilayer/composite substrate interface contributing to the total channel charge. Proper surface preparation of the composite substrate (e.g., chemical and plasma treatment) and development of an insulating buffer schemes will be necessary to remove the parasitic charges at the epilayer/Ga$_2$O$_3$ interface.

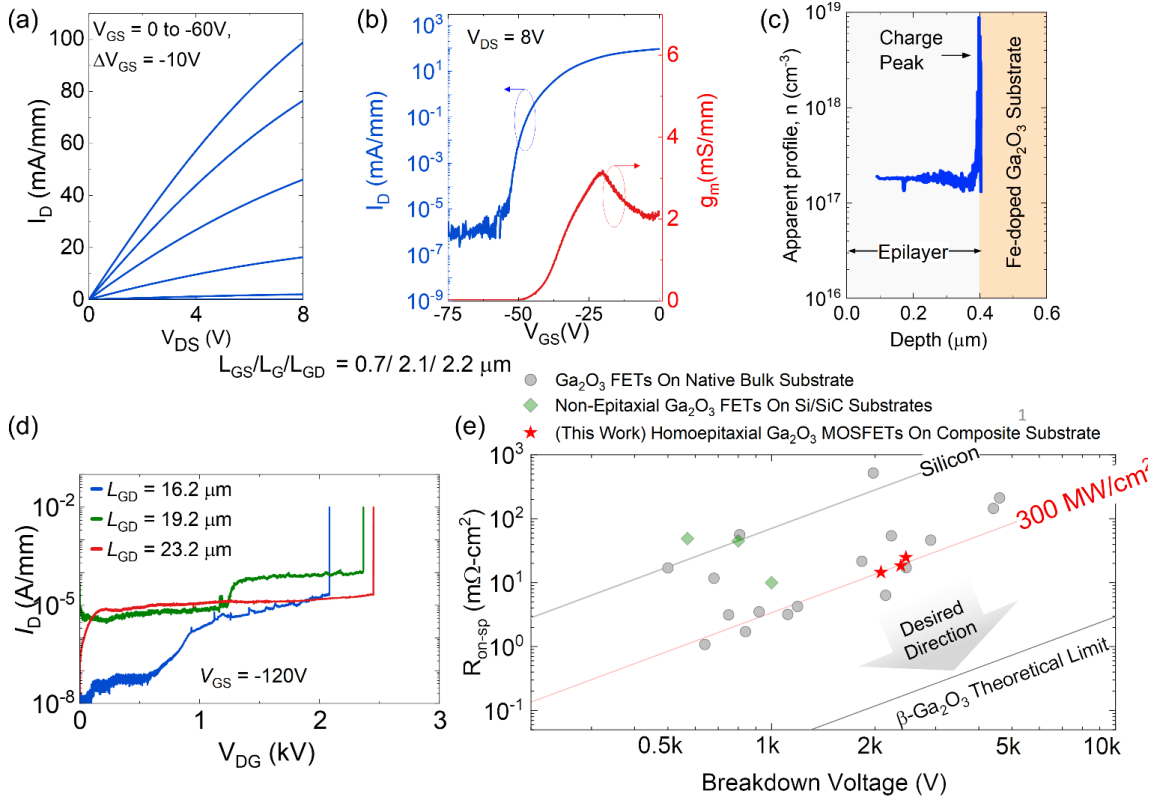

**Figure 3**: (a) DC output and (b) transfer curves of the $Ga_2O_3$ MOSFET fabricated on the composite substrate. (c) Channel charge profile extracted from C-V measurements. (d) Off-state breakdown characteristics of the $Ga_2O_3$ MOSFETs with various $L_{GD}$ values. (e) Benchmarking of the MOSFET fabricated on the composite substrate against homoepitaxial $Ga_2O_3$ FETs[36,37,46–49,38–45] and devices fabricated on other bonded substrates[50–52] in literature.

Device breakdown measurements were performed on the large $L_{GD}$ devices with the wafer submerged in a dielectric liquid (Fluorinert FC-40). A large negative gate bias of -120V was applied during breakdown measurements to minimize bulk-related source-to-drain leakage. The breakdown voltages ($V_{BR}$) increased from 2.08 kV to 2.45 kV as the $L_{GD}$ was scaled from 16.2 μm to 23.2 μm (as shown in Figure 3(d). The average breakdown field was around ~1.3 MV/cm, which is promising given that the devices did not employ any field plates to lower the peak electric field. The specific ON resistance ($R_{on-sp}$; normalized with respect to the device area) values were 14.5, 18.4 and 24.8 mΩ·cm$^2$ for devices with $L_{GD}$ of 16.2, 19.2 and 23.2 μm, respectively. The respective power figure of merit (PFOM[53]; $V_{BR}^2/R_{on-sp}$) of the devices were 295 MW/cm$^2$ ($V_{BR}$=2.08 kV), 303 MW/cm$^2$ ($V_{BR}$=2.37 kV) and 242 MW/cm$^2$ ($V_{BR}$=2.45 kV). These are the highest $V_{BR}$ and PFOM values ever reported for $Ga_2O_3$ transistors fabricated on heterogeneously integrated substrates.[50–52] **Figure 3 (e)** benchmarks the $R_{on-sp}$-$V_{BR}$ performance of the MOSFETs against values reported in literature. With a PFOM of ~300 MW/cm$^2$, these devices are better than most state-of-the-art homoepitaxial $Ga_2O_3$ devices fabricated on native $Ga_2O_3$ substrates and are significantly better than "transferred and non-epitaxial" $Ga_2O_3$ transistors on SiC substrates.[50–52] The electrical performance is not compromised by fabricating devices on the composite substrate when using the low temperature device processing scheme. Even though this is the first demonstration of "epitaxially grown" $Ga_2O_3$ MOSFETs fabricated on a composite substrate, the devices show promising OFF-state voltage blocking capabilities up to 2.45 kV suitable for power electronics applications.

## 2.4. DEVICE THERMAL CHARACTERIZATION

Nano-particle assisted Raman thermometry[54,55] was used to perform *in situ* channel temperature measurement of the MOSFET structures. Anatase $TiO_2$ nanoparticles of 99.98% purity were deposited on the devices to serve as surface temperature probes. The Stokes Raman peak shift of the $E_g$ phonon mode was monitored during device operation to estimate the channel temperature rise. Measurements were taken on nanoparticles close to the drain side edge of the gate, where the channel peak temperature is expected to occur due to electric field and Joule heat concentrations. Measurements were performed on devices with different dimensions ($L_{GD}$ of ~2.2, 28, and 55 μm) fabricated on both the composite wafer and a native $Ga_2O_3$ substrate. **Figure 4** shows the steady-state temperature rise ($\Delta T$) as a function of power density and the corresponding heat flux values. When comparing the temperature rise for devices with different $L_{GD}$, it is important to consider the heat flux since the area where Joule heating occurs is changing. Therefore, a device with a larger $L_{GD}$ will experience a lower temperature rise for a given power density. It should be noted that the devices on the composite and native substrates exhibited similar power densities for particular drain voltages ($V_{ds}$). Due to the enhanced heat transfer performance of the composite substrate, a significant reduction in channel temperature rise (up to a 2.4× reduction) was observed for devices (especially those with larger $L_{GD}$) operating under a power density of 2.63 W/mm.

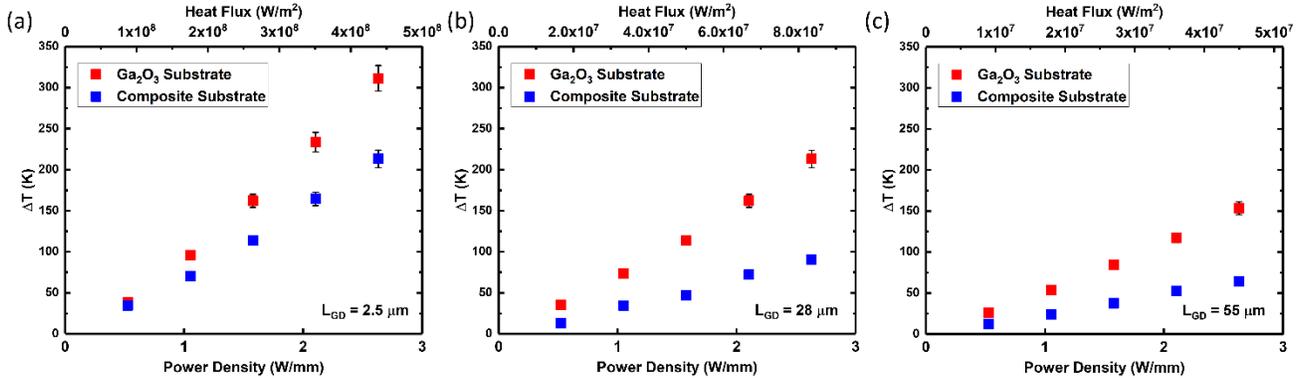

**Figure 4:** Steady-state channel temperature rise of the MOSFETs fabricated on the composite substrate and a bulk $Ga_2O_3$ wafer. Devices with different $L_{GD}$ were tested. (a) $L_{GD}$=2.5 μm, (b) $L_{GD}$=28 μm, and (c) $L_{GD}$=55 μm.

The transient channel temperature rise of the devices was characterized to understand the cooling effectiveness of the composite substrate under high frequency switching operation. A transient Raman thermometry setup (details can be found in the experimental section) was used to monitor the channel temperature rise in response to a square electrical power pulse with a temporal resolution of 25 μs.[54,55] As shown in **Figure 5**, the early-stage temperature rise (<100 μs) is similar between the homoepitaxial and composite substrate; this is because of the low thermal diffusivity (or slow transient thermal response) of $Ga_2O_3$, causing the heating to be restricted within the ~34 μm thick $Ga_2O_3$ layer for both devices fabricated on the composite wafer and a native substrate. A lower temperature rise is observed for the devices on the composite wafer only after 100 μs, from which the high thermal conductivity of 4H-SiC contributes to spreading the heat away from the device active region[3]. Data in **Figure 5** indicate that the channel temperature of the devices on the composite substrate reaches steady-state after 200-300 μs while the temperatures of the devices on a $Ga_2O_3$ substrate continues to increase. In contrast to previously developed diamond integration methods for GaN high electron mobility transistors (HEMTs)[18,56–59], the low thermal diffusivity of $Ga_2O_3$ renders a more in-depth thermal design process required for the development of $Ga_2O_3$ devices on a composite wafer,. For our current design, the thickness of the $Ga_2O_3$ layer of the composite substrate is far larger than ~10 μm (recommended by Chatterjee *et al.*[12]), which is necessary to reduce the device thermal resistance less than that of a GaN-on-Si power switch. Therefore, 3D modeling was performed in the next section to further discuss the transient thermal response and its implications on design optimization.

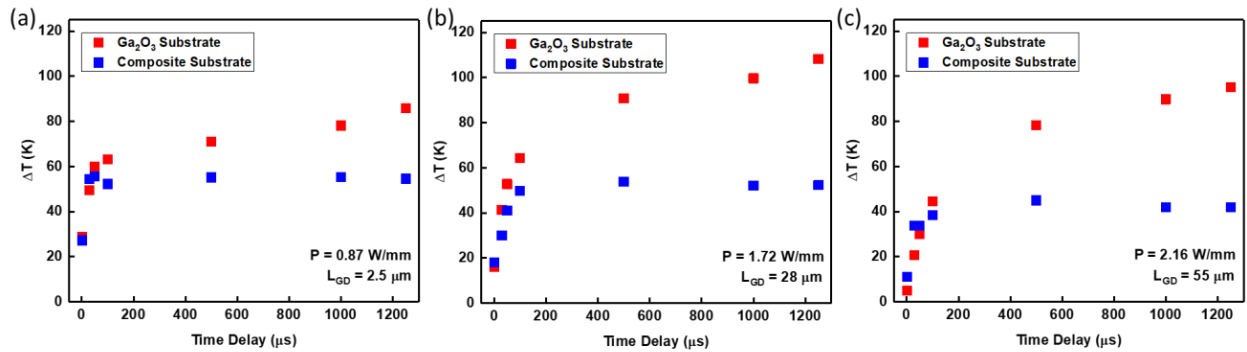

**Figure 5:** The transient channel temperature rise of MOSFETs fabricated on native ($Ga_2O_3$) and composite substrates. Devices with different $L_{GD}$ were tested. (a) $L_{GD}$=2.5 μm, (b) $L_{GD}$=28 μm, and (c) $L_{GD}$=55 μm.

**2.5. MODELING AND DESIGN OPTIMIZATION**

In order to verify the results of the nanoparticle-assisted Raman thermometry measurements, a 3D transient electro-thermal model was constructed.[12,19] The (010) $Ga_2O_3$ layer was modeled to be 34.6 μm based on the SEM results and a directional and temperature dependent thermal conductivity was employed from values published by Guo et al.[6] The 4H-SiC substrate layer was modeled to be 500 μm thick, and a temperature dependent thermal conductivity was adopted from Wei et al.[10] An effective TBR of 47.1 $m^2$K/GW[15] was applied at the $Ga_2O_3$/SiC interface based on SSTR measurement results. First, the device detailed in section 2.2 ($L_{GD}$~2.5 μm) was modeled under steady-state conditions matching the operating conditions used in the nanoparticle Raman experiments listed in **Figure 4 (a)**. **Figure 6** shows the modeling results, which are in excellent agreement with the temperatures measured via the Raman thermometry experiments. The composite wafer is shown to reduce the device peak temperature during operation; however, the thermal performance can be further enhanced by reducing the $Ga_2O_3$ layer thickness, improving phonon transport across the interface, and using a higher thermal conductivity substrate instead of 4H-SiC. Therefore, a second model was built to assess the theoretical performance limit by using such "ideal" composite wafer. This ideal model assumes a reduced $Ga_2O_3$ thickness of 1 μm (which consists of 200 nm from ion-cutting[29], 300 nm from a back barrier, and 500 nm for the channel) and a single crystalline diamond substrate[60]. An effective thermal boundary resistance of 7.8 $m^2$K/GW was assumed between the $Ga_2O_3$ and the diamond substrate based on the TBC for using 10 nm $Al_2O_3$ as the bonding interlayer.[29] This ideal case showed a significant reduction in channel temperature rise (~10×) as compared to the current composite wafer design, suggesting the potential for further enhancement in the cooling performance with optimization.

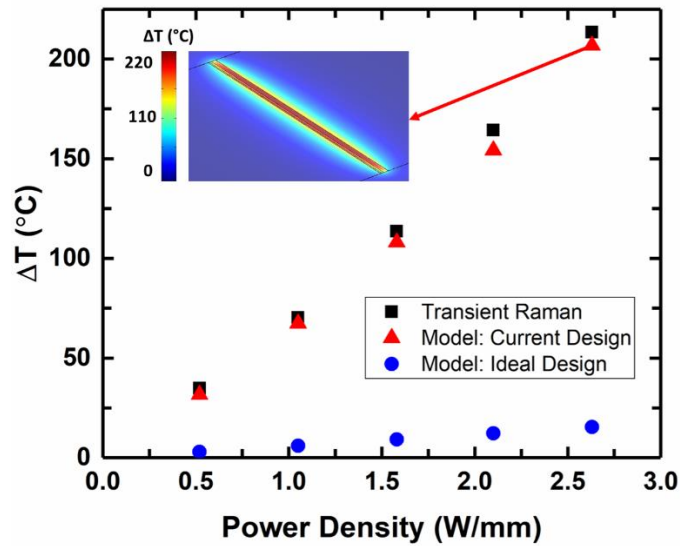

**Figure 6**: Comparison of results (channel temperature rise) from the Raman experiments and simulation for the fabricated $Ga_2O_3$-on-SiC MOSFET and an "ideal" device.

As previously mentioned, a composite wafer must be designed so that high cooling performance is offered under high frequency switching operation. Transient thermal models for both the current design and the ideal case were constructed, with a device ON (power) square pulse of a 1.3 ms period and a 10% duty cycle (to match the transient Raman thermometry experiments). **Figure 7** shows the normalized transient response for the measured and simulated responses for the current design and the ideal case. A power density of 0.87 W/mm was used in this study for both simulation and experiments. Due to the relatively large thickness (~35 μm) of the $Ga_2O_3$ layer, the current design took ~300 μs to reach a quasi-steady state temperature, while this took only ~4 μs for the ideal case. In other words, the current design only offers its full cooling performance for switching frequencies less than ~3 kHz, while the ideal case is effective for frequencies up to ~250 kHz. This switching frequency limit can be further increased by the implementation of top-side cooling solutions such as a diamond passivation overlayer[61] and flip-chipping.[62]

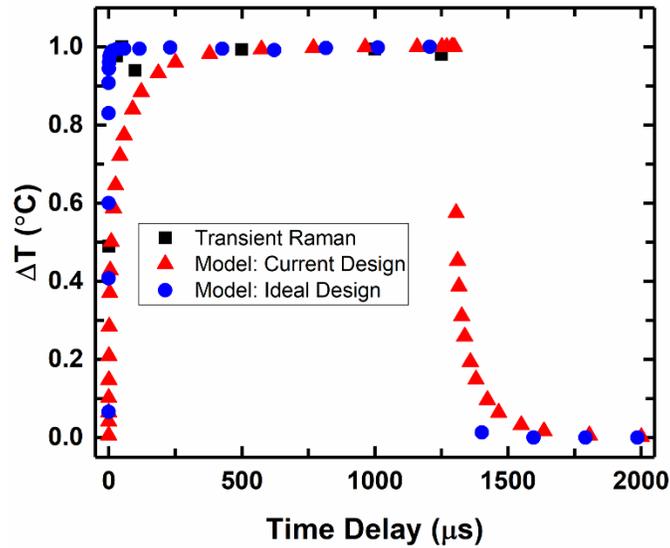

**Figure 7**: Transient thermal response for the current Ga$_2$O$_3$-on-SiC device (measured via Raman thermometry and simulated) and an "ideal" device. The temperature is normalized based on their respective quasi steady-state temperatures (~55°C for the current design and ~6°C for the ideal design).

A recent computational study[15] has predicted that practical multi-finger devices would experience significantly aggravated self-heating (a 4× higher channel temperature than single-finger Ga$_2$O$_3$ devices under identical power density conditions) due to the thermal cross-talk[16] among adjacent current channels. This trend has been experimentally confirmed by an experimental study[17] on multi-channel Ga$_2$O$_3$ FinFETs. Therefore, multi(6)-finger device structures were simulated for both the current Ga$_2$O$_3$/SiC composite wafer and an ideal case as detailed earlier. Further details of the electrically aware thermal model can be found in our previous work.[62] In **Figure 8**, the temperature results can be found for both aforementioned single- and multi (6)-finger Ga$_2$O$_3$ cases, in addition to that for a commercial multi-finger GaN-on-SiC device (details of the device geometry can be found in [62]). Due to the thermal cross-talk between the channels, the temperature rise is greater than that of a single channel device (by comparing with results in **Figures 6** and **8**). A significant reduction in the channel temperature of ~8× is seen in the ideal multi-finger case, giving promise to lower the device thermal resistance below that for today's commercial GaN-on-SiC transistors.[62]

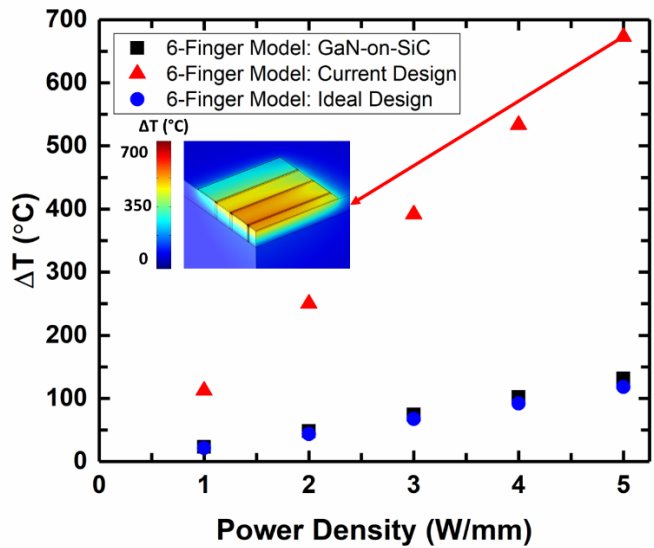

**Figure 8**. Comparison of the real and ideal 6-finger devices temperature rise.

## 3. CONCLUSION

This work reports the fabrication of ultra-wide bandgap $Ga_2O_3$ power MOSFETs on a $Ga_2O_3$/4H-SiC composite wafer with simultaneous enhancement in the electrical and thermal performance. Low temperature (≤600°C) epitaxy and device fabrication processes were developed to preserve the structural integrity of the composite substrate. This enabled the first realization of (010)-oriented "homoepitaxial" $Ga_2O_3$ MOSFETs fabricated on a $Ga_2O_3$/4H-SiC composite substrate. The epitaxial $Ga_2O_3$-on-SiC MOSFETs exhibit a record high $V_{BR}$ (of up to 2.45 kV) and PFOM (~300 MW/cm$^2$), both of which are highest among $Ga_2O_3$ FETs constructed on an heterogeneously integrated substrate to date. Under DC operation, a significant reduction in the channel temperature was achieved for the MOSFETs fabricated on the composite wafer as compared to devices homoepitaxially grown on a native $Ga_2O_3$ substrate. The experimentally measured temperature rise was validated by a 3D FEA electro-thermal model. Transient thermal analysis suggests that the cooling performance of an unoptimized composite wafer will be limited under high frequency switching operation. The theoretical cooling limit of using a hypothetical $Ga_2O_3$/diamond composite substrate with ideal heat transfer performance was assessed by modeling. A 10× improvement in thermal performance can be achieved by reducing the $Ga_2O_3$ layer thickness (to ~1 μm) and lowering the TBR at the $Ga_2O_3$/diamond interface (~7.8 m$^2$K/GW). This work provides key guidelines for the fabrication and realization of high-power UWBG devices on a composite wafer that will enable to surpass the thermal limit of next-generation $Ga_2O_3$ power electronics.

## METHODS

**FIB/scanning electron microscopy**

Plan view and cross-sectional scanning electron microscope (SEM) imaging was performed using a Helios NanoLab$^{TM}$ 650 Dual Beam system with focused ion beam (FIB) milling capabilities with samples under vacuum. For plan view imaging, the samples were imaged with the electron beam perpendicular to the sample surface at low-currents with acceleration voltages of 2 - 5 kV. For cross-sectional imaging, Ga ion current milling with a Pt protective capping layer was used to create craters up to 50 μm deep. Ion current levels of up to 1 nA were used. For charge dissipation, an Omiprobe$^{TM}$ probing system was used to probe the metal contacts of the isolated $Ga_2O_3$ MOSFETs.

**Scanning transmission electron microscopy:**

Scanning transmission electron microscopy (STEM) samples were prepared via focused ion beam (FIB) milling using a Thermofisher Helios Dual-beam FIB system. To prepare a clean and thin specimen, the surface of the STEM foil was cleaned using low energy ion milling (Fischione Nanomill) operated at 500 eV. High angle annular dark field (HAADF) STEM imaging was performed using a Thermofisher aberration-corrected Titan STEM microscope with probe convergence half angles of 10.03 mrad at an accelerating voltage of 300 kV. The microscope is also equipped with ChemiSTEM Energy dispersive X-ray spectroscopy (EDX) system, which allows for the characterization of the composition of the cross-sectional STEM sample. Five chemical species (Ga, Si, O, C, and N) at the interface were analyzed by the EDX elemental mapping. The bonding layer (including $SiN_x$) total thickness of 40 nm was determined at the interface region. Due to the inhomogeneity of the lattice mismatch between the $Ga_2O_3$ thin films and 4H-SiC substrate, the $SiN_x$ bonding interface was marginally delaminated, resulting in the oxidation layer of 10 nm $SiO_x$ within the $SiN_x$ interfacial region. The elemental profile further demonstrated the distribution of O based on the cross-section STEM-EDX measurements.

**Raman thermometry:**

Raman thermometry is a temperature measurement technique that uses Raman spectroscopy, which employs monochromatic photonic excitation (typically in the visible wavelength regime) to interrogate the energy or frequency of crystal lattice vibration (i.e., phonons). The temperature effect on a phonon can be observed in the Raman spectra through peak position shifts, peak broadening (or linewidth), and changes in the ratio of anti-Stokes/Stokes Raman peak intensity.[63,64] Among these three ways, the peak position-based temperature measurement offers higher measurement sensitivity with low uncertainty, and shorter measurement times. However, this method can lead to inaccuracies due to its sensitivity to both temperature and mechanical stress. Moreover, this measurement technique only provides the depth-averaged temperature information for UWBG semiconductors such as $Ga_2O_3$.

In this study, both steady-state and transient Raman thermometry were performed using a Horiba LabRAM HR Evolution spectrometer with a 532 nm excitation laser. A long working distance 50× objective (NA=0.45) was used in a 180° backscattering configuration. A nanoparticle-assisted Raman thermometry technique[65] was used to measure the surface temperature of the channel region of the $Ga_2O_3$ MOSFETs. Anatase titanium dioxide ($TiO_2$) nanoparticles with 99.98% purity were deposited on the device surface.[65] As the nanoparticle remains in thermal equilibrium with the device surface, the temperature dependent frequency shift of the $E_g$ phonon mode mode was monitored to determine the device channel temperature. Since the nanoparticle can expand freely, the mechanical stress effect on the Raman peak position is negligible and does not affect temperature measurement results. The spatial resolution is determined by the size of the $TiO_2$ nanoparticles (~200 nm).

By augmenting the standard Raman microscope with a function generator, delay generator, trigger switch, and oscilloscope, a setup for transient temperature measurement was constructed.[66] This transient setup uses a lock-in modulation scheme in which electrical and laser pulse trains are synchronized, and the Raman signal accumulates over many periods. A full transient thermal response is constructed by controlling/sweeping the laser pulse delay time ($\tau_{delay}$) along the entire device electrical ($V_{DS}$) pulse width ($\tau_{on}$). An exemplary temporal schematic of the transient measurement is shown in **Figure 9**. In this example, $V_{DS}$ ~35 V (i.e., electric pulse) is applied to achieve a power dissipation level of ~0.5 W while the device is ON. The laser pulse is active at the very end of the electrical pulse. A digital delay generator controls the time delay ($\tau_{delay}$) between the electrical and laser pulses, in order to monitor the transient self-heating behavior of the device in response to a square electrical pulse with a 10% duty cycle. Here, the device pulse width ($\tau_{on}$) is 1.3 ms and the laser pulse width ($\tau_{laser}$) is 25 μs. A temporal resolution up to ~20 ns can be achieved with our experimental setup. The synchronization and operation of the transient measurements were controlled by a LabVIEW program. To maximize the signal-to-noise ratio of the Raman signals, an electron multiplying charge-coupled device (EMCCD) was used during the measurements.

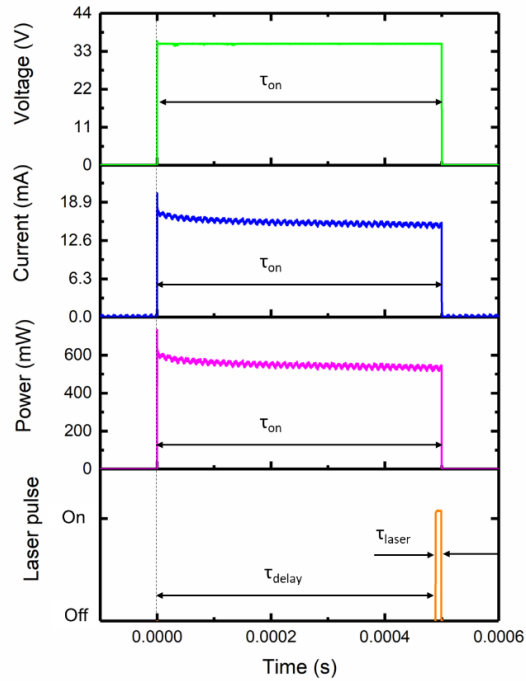

**Figure 9**. A typical synchronized pulsing scheme used during transient Raman thermometry measurements.

**Thermo-physical property measurement:**

Steady-state thermoreflectance (SSTR) is a laser-based pump-probe technique, which is ideal to measure the thermal conductivity of bulk materials.[24] Details of the SSTR setup used in this study is described in our previous work.[15] The pump and probe lasers were focused on the samples using the following microscope objectives: (i) a 2.5× objective (NA = 0.08) with pump and probe laser radii of 19.4 μm and 12.4 μm, respectively, (ii) a 10× objective (NA = 0.25) with pump and probe laser radii of 5 μm and 4.3 μm, respectively. The thermal conductivity of the $Ga_2O_3$ layer of the composite substrate was measured using a 10× objective to confine the probing volume within the $Ga_2O_3$ layer. The TBR at the interface was derived by performing measurement using a 2.5× objective and fitting the data based on the measured thermal conductivity values of $Ga_2O_3$ and 4H-SiC.

Time domain thermoreflectance (TDTR) is an optical pump-probe technique that allows the extraction of thermal properties of thin films based on heat diffusion in response to ultrafast femtosecond laser pulses.[25,26] Details of the TDTR setup used in this study are described in our previous work[15]. The radii of the focused pump and probe beams were characterized using a scanning-slit optical beam profiler, and their radii were determined to be 8.4 μm and 6 μm, respectively. TDTR (and FDTR) measurements were performed on a wedge-shape thinned $Ga_2O_3$ composite substrate to determine the TBR at the $Ga_2O_3$/4H-SiC interface. The thermal boundary conductance between the metal transducer and the $Ga_2O_3$ film as well as the cross-plane thermal conductivity of $Ga_2O_3$ were simultaneously determined by TDTR measurements. To fit the acquired data, literature values for the thermal conductivity and volumetric heat capacity of Au[67] as well as the volumetric heat capacities of $Ga_2O_3$[68] and 4H-SiC[69] were assumed. The uncertainties in data were calculated based on 95% confidence bounds resulting from multiple measurements; error propagation associated with ±2 nm uncertainty in the metal transducer thickness have also been accounted for.

Frequency domain thermoreflectance (FDTR) is an optical pump-probe technique that measures material thermal properties based on fitting the phase of the thermal wave over a range of modulation frequencies.[27,28] Details of the FDTR setup used in this study are described in our previous work.[15] The radius of the focused pump and probe beams were 13.4 μm and 13.1 μm, respectively. Material properties used to post-process the FDTR raw data were identical to those used in the analytical model for TDTR experiments. The $Ga_2O_3$ thermal conductivity (from TDTR measurements) and the 4H-SiC thermal conductivity (determined by SSTR) of the wedge-shape thinned composite substrate were used as known parameters to extract the TBR between the $Ga_2O_3$ layer and 4H-SiC substrate.

**Device modeling**

$Ga_2O_3$ device models were constructed using a 3D electro-thermal modeling scheme that has been demonstrated in our previous work.[12,19] A 2D electro-hydrodynamic model that adopts carrier mobility and Ohmic contact resistance determined from experiments is created so that it reproduces the device output/transfer characteristics. This electrical model calculates the internal heat generation profile as a function of electrical bias. The 2D Joule heat distribution is projected along the channel width so that a 3D volumetric heat generation profile is obtained. This 3D heat generation profile is imported into a 3D finite element transient thermal model. The $Ga_2O_3$ and 4H-SiC thermal conductivities determined from experiments are employed in the thermal model. Interfacial phonon transport across the $Ga_2O_3$/4H-SiC interface is captured in the model by adopting the TBR values determined by experiments.

**AUTHOR INFORMATION**

**Corresponding Authors**

* E-mail: sukwon.choi@psu.edu and sriramkrishnamoorthy@ucsb.edu.

**Author Contributions**

Y.S. and A.B. have equally contributed to this work as co-first authors. S.C. and S.K. are co-corresponding authors of this work. Y.S. performed thermo-physical property measurement and led writing the manuscript. A.B. and S.R. fabricated the $Ga_2O_3$ MOSFET. A.B. contributed to writing the manuscript. A.K. performed optical thermometry. D.S. created the device thermal model and performed computational analysis. H-L.H. and J.H. performed electron microscopy studies. C.M. and J.H.L. fabricated the composite substrate. S.K. directed device fabrication and electrical characterization efforts in this work. S.C. managed the entire project, designed experiments and computational studies, and wrote the manuscript. All authors have given approval to the final version of the manuscript.

**Acknowledgments**

Funding for efforts by the Y.S. and S.C, was provided by NSF (CBET-1934482, Program Director: Dr. Ying Sun) and the Air Force Office of Scientific Research (AFOSR) Young Investigator Program (FA9550-17-1-0141, Program Officers: Dr. Brett Pokines and Dr. Michael Kendra, also monitored by Dr. Kenneth Goretta). A.B. and S.K. were supported by the II-VI Foundation's Block Gift Program and AFOSR (FA9550-21-0078, Program Officer: Dr. Ali Sayir). This work was performed in part at the Utah Nanofab sponsored by the College of Engineering and the Office of the Vice President for Research. A. B. and S. K. thank Dr. Randy Polson at the Electron Microscopy and Surface Analysis Lab (EMSAL), University of Utah for the FIB/SEM measurements and analysis used in this work. H.H. and J.H. acknowledge support by the AFOSR (FA9550-18-1-0479, Program Officer: Dr. Ali Sayir). Electron microscopy was performed in the Center for Electron Microscopy and Analysis (CEMAS) at The Ohio State University.